\documentclass[aps, pra,twocolumn,showpacs,preprintnumbers,amsmath,amssymb]{revtex4}
\usepackage{graphicx}
\usepackage{dcolumn}
\usepackage{bm}
\usepackage{psfrag}
\usepackage{epsfig}
\usepackage{amsmath}
\usepackage{amssymb}
\usepackage{color}
\usepackage[FIGTOPCAP,raggedright,nooneline]{subfigure}
\newcommand{\bra}[1]{\langle #1 |}
\newcommand{\ket}[1]{| #1 \rangle}

\newcommand{\id}{{\sf 1 \hspace{-0.3ex} \rule{0.1ex}{1.52ex}\rule[-.01ex]{0.3ex}{0.1ex}}}
\newcommand{\ignore}[1]{}

\newcommand{\I}{\ensuremath{\mathrm{i}}}
\begin{document}

\title{Reaching the optomechanical strong coupling regime with a single atom in a cavity}

\author{Lukas Neumeier$^1$}
\email{lukas-neumeier@gmx.de}
\author{Tracy E. Northup$^2$}
\author{Darrick E. Chang$^{1,3}$}
\email{Darrick.Chang@icfo.eu}
\affiliation{$^1$ICFO-Institut de Ciencies Fotoniques, The Barcelona Institute of Science and Technology, 08860 Castelldefels, Barcelona, Spain}
\affiliation{$^2$Institute for Experimental Physics, University of Innsbruck, A-6020 Innsbruck, Austria}
\affiliation{$^3$ICREA-Instituci\'{o} Catalana de Recerca i Estudis Avan\c{c}ats, 08015 Barcelona, Spain}

\date{\today}

\begin{abstract}
A major goal within the field of optomechanics is to achieve the single-photon strong coupling regime, wherein even a mechanical displacement as small as the zero-point uncertainty is enough to shift an optical cavity resonance by more than its linewidth. This goal is difficult, however, due to the small zero-point motion of conventional mechanical systems. Here, we show that an atom trapped in and coupled to a cavity constitutes an attractive platform for realizing this regime. In particular, while many experiments focus on achieving strong coupling between a photon and the atomic internal degree of freedom, this same resource also naturally enables one to obtain optomechanical strong coupling, in combination with the low mass of an atom and the isolation of its motion from a thermal environment. As an example, we show that optomechanically-induced photon blockade can be realized in realistic setups, and provide signatures of how this effect can be distinguished from the conventional Jaynes-Cummings blockade associated with the two-level nature of the atomic transition.
\end{abstract}
\maketitle

Spectacular advances in optomechanics now allow quantum control over the interaction between photons and phonons \cite{aspel}. In experiments thus far, a large classical field drives the system, around which the light-motion interaction can be linearized. This linear interaction enables many applications, ranging from laser cooling of the motion to its ground state \cite{oscarpainerlasercooling, sidebandcooling}, squeezed light generation \cite{sqeezedlight1,cindy}, sensing \cite{sensing1,sensing2}, microwave-to-optical conversion \cite{microw,microw2}, and non-reciprocal optical devices \cite{nonrec,nonrec2}. However, the interaction between photons and phonons is intrinsically nonlinear. These nonlinear effects can be observed in strongly driven systems (e.g., self-sustained oscillations \cite{marg1,marg3, osc,osc2}. A particularly interesting limit to consider is when this nonlinearity manifests itself at the level of single-photon inputs \cite{rabl,marg2}, leading to the ability to generate highly non-Gaussian states.
To reach this regime, a zero-point mechanical displacement should shift the frequency of the optical resonator by an amount comparable to its linewidth, which is difficult due to the large mass of conventional mechanical elements and the implied small zero-point motion. While a number of schemes have been proposed to attain optomechanical strong coupling \cite{oth,oth1,oth2}, this regime has yet to be experimentally demonstrated. Thus, finding a platform where this regime can be explored constitutes a major goal.

Separately, much work focuses on coupling single neutral atoms \cite{photonblockadekimble,reiserer,rauschen,rooting} or ions \cite{ion1,ion2,ion3,ion4,ion5,ion6,ion7} to high-finesse cavities. Here, the primary motivation is to exploit the two-level atomic structure to generate non-classical fields. However, in this Letter, we argue that such systems also constitute a natural platform to reach the strongly interacting limit of optomechanics. In particular, experiments \cite{photonblockadekimble,reiserer,prl1,prl2} now routinely reach the strong coupling regime of cavity QED, wherein an atom maximally coupled to the cavity (in an anti-node) shifts the bare cavity frequency by more than a linewidth. Moving the atom by a quarter-wavelength to a node eliminates this shift. Thus, a zero-point motion on the order of a fractional wavelength is sufficient to attain optomechanical strong coupling, which is easily achievable given the light single-atom mass.
In addition, the motion of the atom is effectively isolated from a thermal environment, providing the coherence times necessary to produce interesting quantum behavior. As a specific example, we show theoretically that one can observe optomechanically induced photon blockade \cite{rabl} in realistic cavity QED setups, where a non-classical anti-bunched field is produced as the system is unable to transmit more than a single photon at a time. We also describe how this optomechanical behavior can be clearly distinguished from, and dominate over, the usual anti-bunching associated with the two-level nature of the atom. The explicit use of the strong coupling regime of cavity QED to attain novel regimes of optomechanics, and the examination of the resulting non-classical statistics of the outgoing field, distinguish the present work from previous experiments that explored optomechanical effects with atomic ensembles in cavities \cite{nat1,prl1,prl2}.

\section{Optomechanical photon blockade}
\begin{figure}
     \centering
\includegraphics[width=0.5\textwidth]{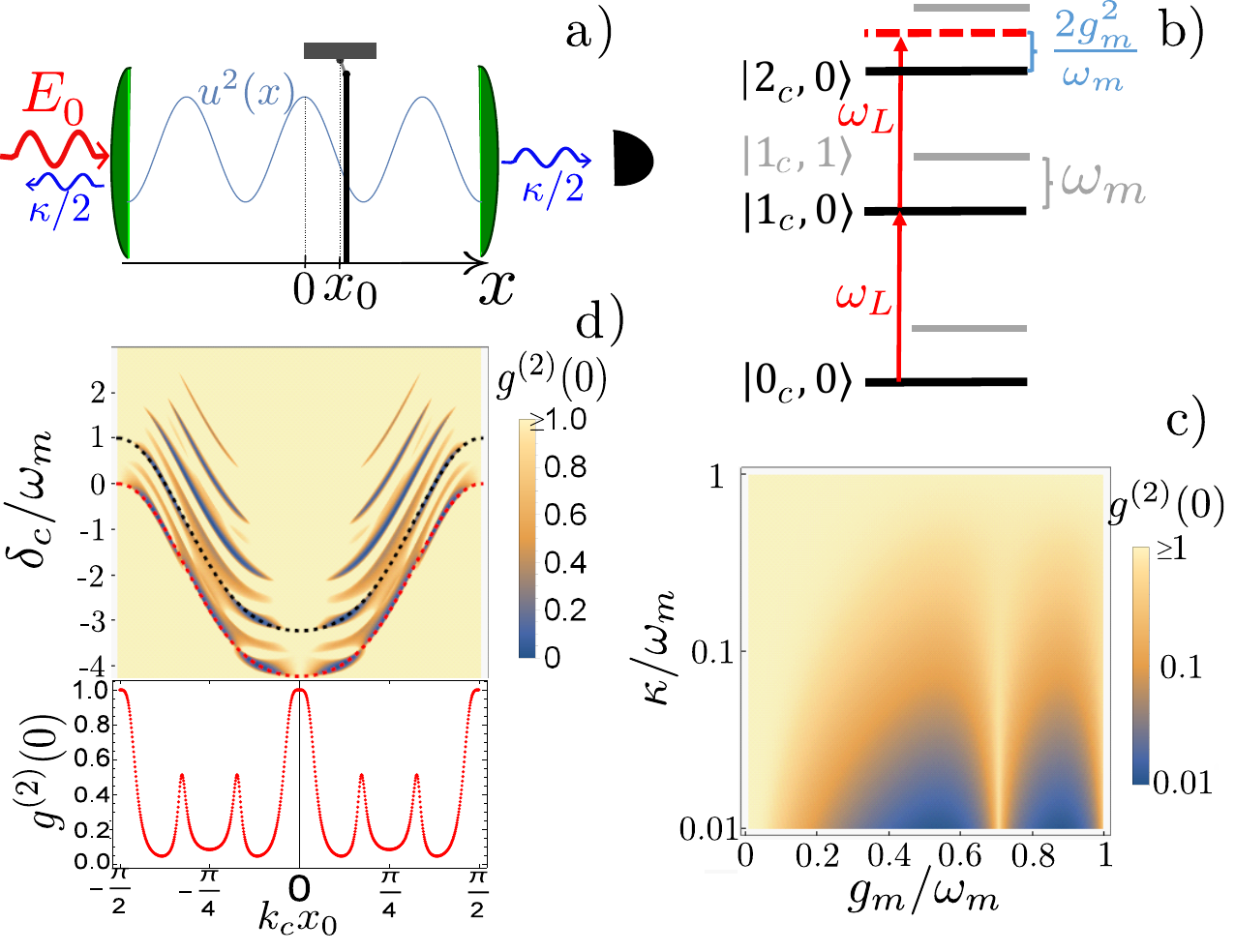}
\caption {Optomechanical photon blockade. \textbf{a)} A membrane with equilibrium position $x_0$ inside a cavity with intensity mode profile $u^2(x)$, which is driven with number flux $E_0^2$ from the left. Each mirror has a decay rate of $\kappa/2$. The photons are measured on the transmitting side of the cavity (right). \textbf{b)} Spectrum of the optomechanical Hamiltonian $H_\mathrm{op}$ for $E_0 \to 0$. Here, $\ket{n,m}$ denotes the state with $n$ photons and $m$ phonons. In this diagram, we focus on transitions involving states with $m=0$ phonons (black lines), while other states ($m=1$ shown here) are denoted by gray lines. A laser with frequency $\omega_L$, which is resonant with the transition $ \ket{0_c,0} \rightarrow \ket{1_c,0}$ (the zero-phonon line), cannot resonantly excite a second photon $\ket{2_c,0}$ as optomechanical interactions shift the relative energy of this state by an amount $2g_m^2/\omega_m$.
\textbf{c)} Normalized second-order correlation function of the transmitted field, $g^{(2)}(0)$, as a function of $g_m/\omega_m$ and $\kappa/\omega_m$. \textbf{d)}Top: $g^{(2)}(0)$ as a function of equilibrium position $x_0$ and detuning from the empty cavity $\delta_c = \omega_L - \omega_c$, normalized by the trap frequency $\omega_m$. The mechanical system is coupled to an intensity mode profile $u^2(x)=\cos^2 (k_c x)$, where $k_c$ is the wavevector of the cavity mode. The dashed red/black lines denote a detuning where the cavity is resonantly driven on the zero phonon line (ZPL)/first phonon sideband, respectively. Bottom: value of $g^{(2)}(0)$ along the ZPL.  The parameters chosen for Fig.~1d) are $g_\mathrm{m0}=2 \pi \times 0.16 \,\mathrm{MHz}$, $\kappa= 2 \pi \times 0.02 \,\mathrm{MHz}$, $\omega_m = 2 \pi \times 0.2 \,\mathrm{MHz}.$
}
\label{fig1}
\end{figure}

We begin by reviewing the phenomenon of photon blockade in a conventional optomechanical system.
We focus on the system shown in Fig.~\ref{fig1}a), where a mechanical element such as a trapped particle \cite{aspelev,barker1,barker2,aspel2, lukas} or membrane \cite{jack} can be positioned arbitrarily, and couples to a single standing-wave optical mode of a Fabry-Perot cavity. For small displacements of the mechanical degree of freedom around the equilibrium position $x_0$, the cavity frequency is given by $ \omega_c(x) \approx \omega_c(x_0) + \omega_c'(x_0) (x-x_0)$. The total Hamiltonian of the system, including a coherent external driving field, is given in a frame rotating with the laser frequency $\omega_L$ by
\begin{align}\label{int}
 H_{\mathrm{op}} =  & \quad \nonumber \omega_m b^{\dagger}b -(\omega_L - \omega_c(x_0)+ \I \frac{\kappa}{2})  a^\dagger a \\ & + g_m (b+b^{\dagger})a^{\dagger}a + \sqrt{\frac{\kappa}{2}}E_0(a^\dag + a).
\end{align}
Here, $\omega_m$ is the frequency of the vibrational mode, and $a$ and $b$ denote the photon and phonon annihilation operators, respectively. The quantity $\omega_L - \omega_c(x_0)$ is the detuning between laser frequency $\omega_L$ and the cavity frequency $\omega_c(x_0)$ when the mechanical system lies at its equilibrium position. Each cavity mirror has a decay rate of $\kappa/2$ into outgoing radiation, while the left side also serves as the source of injection of a coherent state into the cavity with photon number flux $E_0^2$.
The position-dependent cavity shift described previously has been re-written in terms of phonon operators as $\omega_c'(x_0)(x-x_0)=g_m(b+b^{\dagger})$ where  $g_m = \omega_c'(x_0) x_\mathrm{zp}$ is the single photon-phonon coupling strength and $x_\mathrm{zp}= \sqrt{\hbar/(2m_\mathrm{eff}\omega_m)}$ is the zero-point motional uncertainty ($m_\mathrm{eff}$ being the effective mass).
The cubic interaction term $(b+b^{\dagger})a^{\dagger}a$ gives rise to nonlinear equations of motion, but quantum signatures have not been observed, as the best ratio of coupling strength to linewidth so far is $g_m/\kappa\sim 10^{-2}$ \cite{oth1, dingens}. Thus, current experiments remain in the so-called optomechanical weak coupling regime, where many photons inside the optical mode are required to see an appreciable interaction, and allowing for linearization around the strong classical cavity field. However, here we will focus on the regime where this linearization breaks down and the nonlinear nature of the optomechanical coupling manifests itself via photon coincidence measurements \cite{rabl}.

To quantify the optomechanical nonlinearity we change into a displaced oscillator representation, which diagonalizes $H_\mathrm{op}$ in the limit of weak driving \cite{rabl}. The eigenvalues as $E_0 \to 0$ can then be written as
$
 E_\mathrm{n,m} = m \omega_m + n \omega_c(x_0)  - \frac{g_m^2}{\omega_m}  n^2
$
and correspond to the (displaced) eigenstates $\ket{n,m}$. The spectrum is shown in Fig.~\ref{fig1}b). If the laser frequency is resonant with the transition $\ket{0_c,0} \rightarrow \ket{1_c,0}$ (zero phonon line $\equiv$ ZPL) then the transition for the second photon
is off resonant from the transition $\ket{1_c,0} \to \ket{2_c,0}$ by an amount $E_{2,0} - 2 E_{1,0} = - 2 g_m^2/\omega_m$.
In order to have a substantial effect, this anharmonicity should be resolvable, $g_m^2/\omega_m\gtrsim\kappa$, and furthermore, one should operate in the sideband resolved regime $\omega_m\gtrsim\kappa$ so that transitions to other motional states, e.g., the first phonon sideband $\ket{0_c,0}\rightarrow \ket{1_c,1}$ are suppressed.
  These requirements for antibunching can also be observed in Fig.~\ref{fig1}c), where we have plotted the second-order correlation function $g^{(2)}(0$) of the transmitted field given a weak coherent state input for different values of  $\kappa$ and $g_m$, taking the laser frequency $\omega_L$ as being resonant with the ZPL (see Appendix A for details of the calculation). A value of $g^{(2)}(0)<1$ indicates non-classical antibunching, and a minimum value occurs around around $g_m \approx 0.5\omega_m$, which for well-resolved sidebands decreases as $g^{(2)}(0)\approx 20(\kappa/\omega_m)^2$. One also sees that increasing the ratio $g_m/\omega_m$ further does not improve the amount of antibunching, due to the possibility of resonantly coupling to other excited states. For example, at $g_m/\omega_m \approx 1/\sqrt{2}$, the reduced antibunching arises as a second photon can resonantly excite the state $\ket{2_c,1}$, since $E_{2,0}-2E_{1,0}=-\omega_m$.

While mathematically the degree of antibunching is determined by the parameters $g_m,\omega_m,\kappa$, it will also be helpful to ``visualize'' how the antibunching changes as the equilibrium position $x_0$ is scanned from a cavity anti-node to node, to provide a useful comparison with atoms later. For a weak dielectric perturbation such as a thin membrane, intuitively one expects that the variation in the cavity frequency follows the intensity profile of the standing wave itself, $\delta\omega_c(x)\propto -\cos^2 (k_c x)$ \cite{opto,darrick}. It follows then that $g_m(x_0)= g_\mathrm{m0} \sin(2k_c x_0)$. In particular, $g_m(x_0)$ vanishes at a node or anti-node, and reaches the maximum possible value of $g_\mathrm{m0}$ halfway between. In Fig.~\ref{fig1}d) we plot $g^{(2)}(0)$  as a function of trapping position $x_0$ and detuning from the empty cavity $\delta_c = \omega_L - \omega_c$ for a mechanical system initially in its ground state. The dashed red line corresponds to a driving laser resonant with the ZPL, which requires the laser frequency to be tuned following the energy eigenvalue $E_\mathrm{1,0}$. In addition to the features along the ZPL, antibunching can also be observed when a motional sideband $\ket{1_c,m}$ is resonantly driven, following the equation $\omega_L=E_\mathrm{1,m}$ (see black dashed curve for m=1). 
Below, we plot
$g^{(2)}(0)$ following the ZPL (red, dashed).
The oscillations in $g^{(2)}(0)$ along the ZPL versus $x_0$ occur as $g_m(x_0)$ sweeps into and away from the optimal values for antibunching (compare with Fig. 1c)).
Here, we have chosen parameters of $g_{m0}=2\pi\times  0.16 \,\mathrm{MHz} $, $\kappa =2 \pi \times 0.02 \,\mathrm{MHz}$ and $\omega_m = 2\pi \times 0.2 \,\mathrm{MHz}$. These do not necessarily correspond to a physically realizable optomechanical system, but allow the interesting features to be observed.
\section{Cavity QED without motion}
\begin{figure}
     \centering
\includegraphics[width=0.5\textwidth]{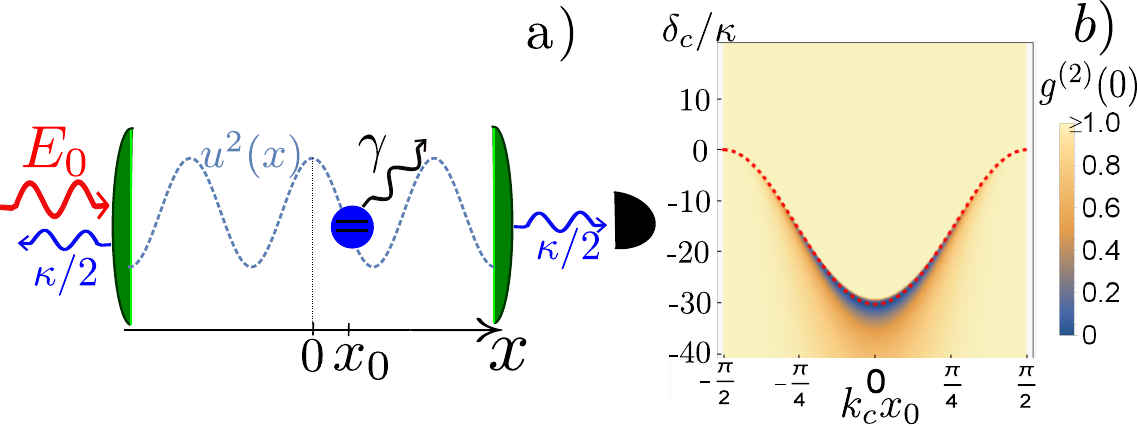}
\caption{Cavity QED without motion. \textbf{a)} Schematic of an atom infinitely tightly trapped inside a cavity mode at position $x_0$. The cavity and atomic excited state decay rates are $\kappa$ and $\gamma$, respectively. \textbf{b)} Second-order correlation function $g^{(2)}(0)$ of the transmitted field, as a function of trapping position $x_0$ and detuning from the empty cavity $\delta_c = \omega_L - \omega_c$, normalized by the cavity linewidth $\kappa$. Here, we restrict ourselves to driving frequencies near the resonance of the photon-like dressed state of the Jaynes-Cummings model.
To generate this plot, we take idealized parameters such that antibunching arising from strong atom-cavity coupling can be easily seen: $\Delta = 3 g_0$, $g_0 = 2 \pi \times 2 \,\mathrm{MHz}$, $\kappa = \gamma =2 \pi \times 0.02 \,\mathrm{MHz}$.
}
\label{fig2}
\end{figure}

We now consider an atom coupled to a cavity mode with amplitude $u(x)=\cos(k_c x)$ (see Fig.~\ref{fig2}a)), which is described by the Jaynes-Cummings (J-C) Hamiltonian \cite{jaynes}. Due to the two-level nature of the atom, the spectrum of the J-C Hamiltonian is nonlinear. We thus study the effect of this nonlinearity on $g^{(2)}(0)$ first without motion (i.e., the atom is infinitely tightly trapped), so that we can later clearly distinguish motional effects. The J-C Hamiltonian, in an interaction picture rotating at $\omega_L$, is given by
\begin{align}\label{ham}
 H_\mathrm{JC} = & \nonumber - (\delta_0+ \I \frac{\gamma}{2}) \sigma_\mathrm{ee} -(\delta_c + \I \frac{\kappa}{2} ) a^\dag a  \\ &+  \sqrt{\frac{\kappa}{2}} E_0 (a+a^\dag)+  g_0 u(x_0)(a^\dag \sigma_\mathrm{ge} + h.c.).
\end{align}
The laser-atom detuning is $\delta_0 = \omega_L-\omega_0$ with $\omega_0$ being the resonance frequency of the atom, while $\sigma_{\mathrm{\alpha\beta}}=\ket{\alpha}\bra{\beta}$, where ${\alpha,\beta}={g,e}$ correspond to combinations of the atomic ground and excited states. As before, $\delta_c=\omega_L-\omega_c$ is the detuning relative to the bare cavity resonance. The atom-cavity coupling strength $g_0 u(x_0)$ depends on the trapping position $x_0$, where $g_0$ is the magnitude of the vacuum Rabi splitting at the anti-node at the cavity waist.
The emission rate of an excited atom into free space is given by $\gamma$.

Ignoring dissipative processes for the moment, the system is block diagonal for $n$ total excitations in the system, with possible states $\ket{g,n},\ket{e,n-1}$. The energy eigenvalues in each block are given by $E_n^{\pm} = n\omega_c +(\pm \sqrt{4 g_0^2 u^2(x_0) n+\Delta^2}+ \Delta)/2$, where $\Delta = \omega_0-\omega_c$. In the following we consider the dispersive regime $\Delta \gg g_0, \kappa, \gamma$, where the single-excitation eigenstates of the J-C Hamiltonian are either mostly atomic ($\ket{\psi_+} \approx \ket{e,0}$) or photonic ($\ket{\psi_-}\approx \ket{g,1}$).
These eigenstates have corresponding eigenenergies $E_1^+ \approx \omega_0 +\frac{g_0^2}{\Delta} u^2(x_0)$ and $E_1^- \approx \omega_c -\frac{g_0^2}{\Delta} u^2(x_0)$, respectively. Here, we focus on the case when the system is driven near resonantly with the photonic eigenstate. In that limit, the atom can approximately be viewed as a classical dielectric that provides a position-dependent cavity shift $\propto \frac{g_0^2}{\Delta}$. However, the two-level nature of the atom provides a residual nonlinearity to excite a second photon, of magnitude $E_2^- - 2 E_1^- \approx 2 (g_0^4/\Delta^3) u^4(x_0)$. Such a nonlinearity results in an anti-bunched transmitted field if it is comparable to the cavity linewidth $\kappa$. In Fig.~\ref{fig2}b) we plot $g^{(2)}(0)$ for $\Delta = 3 g_0$, as a function of atom position $x_0$ and detuning $\delta_c$, for frequencies around the photonic eigenenergy $E_1^{-}$ (dotted line). Here, we have chosen idealized parameters $g_0 = 2 \pi \times 2 \,\mathrm{MHz}$, $\kappa = \gamma =2 \pi \times 0.02 \,\mathrm{MHz}$, which enable the antibunching features to be clearly seen. Without motion, the largest degree of antibunching naturally occurs around the anti-node $(x_0=0)$ and monotonically decreases as one approaches the nodes.
\section{Full model: Cavity QED with motion}
We now include atomic motion into the Jaynes-Cummings Hamiltonian $H = \omega_m b^\dag b + H_\mathrm{JC}$ by treating $x_0 \to x$ as a dynamical variable. We assume that the atom sees an internal-state independent and harmonic trapping potential, which occurs naturally for trapped ions or using magic wavelength traps for neutral atoms \cite{mags}.
\begin{figure}
     \centering
\includegraphics[width=0.5\textwidth]{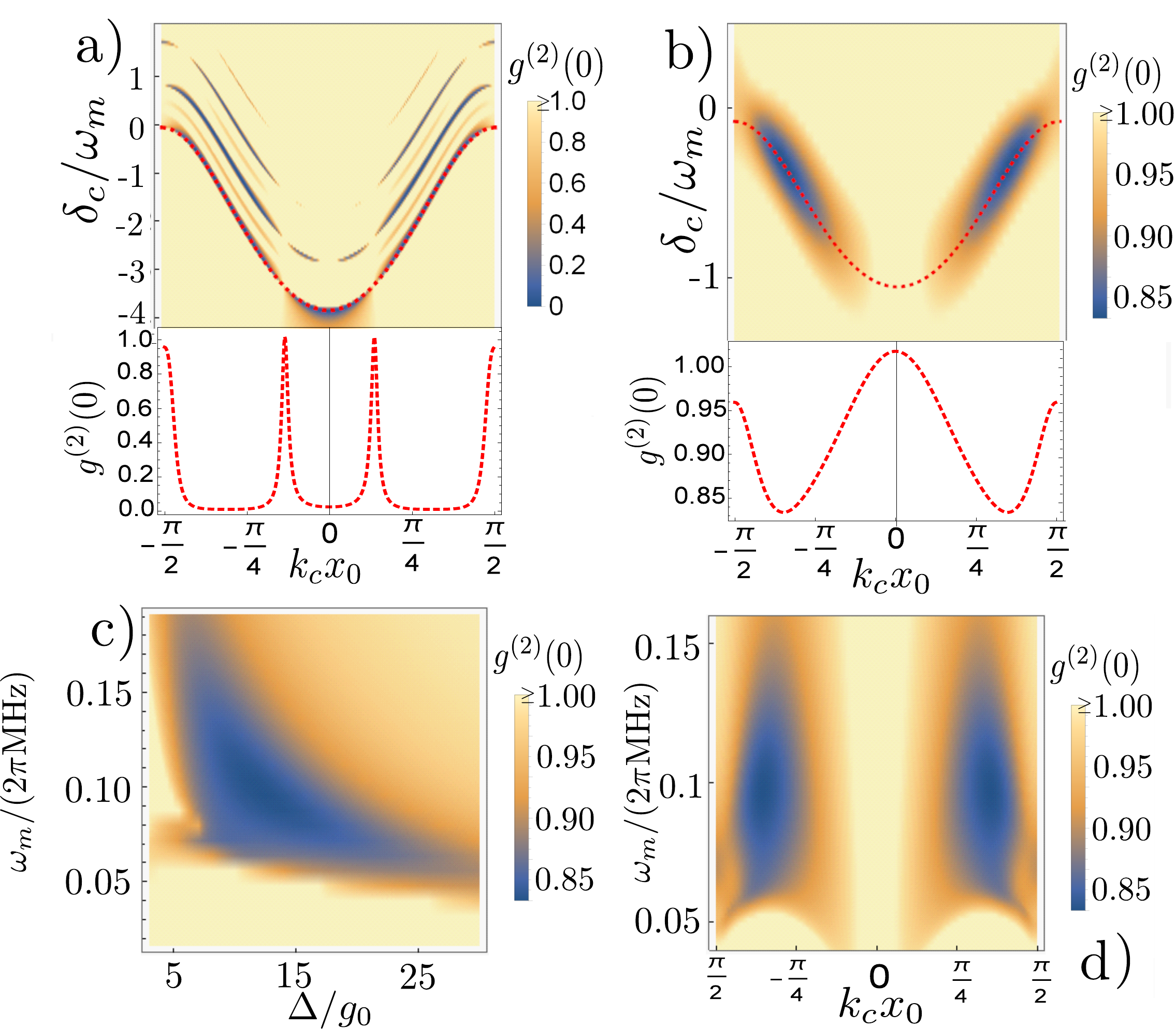}
\caption{J-C model including motion. \textbf{a)}Top: $g^{(2)}(0)$ of the transmitted field versus trapping position $x_0$ and detuning from the empty cavity $\delta_c=\omega_L-\omega_c$, for detunings near the photonic eigenstate and for atom-cavity detuning $\Delta=5g_0$. Here, we use idealized parameters $g_0 = 2 \pi \times 10 \,\mathrm{MHz}$, $\kappa = \gamma =2 \pi \times 0.02 \,\mathrm{MHz}$ , and $\omega_m= 2 \pi \times 0.5 \, \mathrm{MHz}$ so that all of the key features can be clearly observed. Below: $g^{(2)}(0)$  following the ZPL (red, dashed).  \textbf{b)} We plot the same as in Fig.~\ref{fig4}a), but using the parameters for a realistic cavity QED experiment given below. In this figure, we choose $\Delta=12g_0$ and $\omega_m = 2\pi \times 0.1 \,\mathrm{MHz}$.
\textbf{c)} $g^{(2)}(0)$  as  a function of atom-cavity detuning $\Delta$ and trapping frequency $\omega_m$.
\textbf{d)} $g^{(2)}(0)$  as  a function of trapping position $x_0$ and trapping frequency $\omega_m$ for $\Delta = 12g_0$. For Figs.~\ref{fig4}b), \ref{fig4}c) and \ref{fig4}d) we choose parameters $g_0 = 2 \pi \times 1.4 \,\mathrm{MHz}$, $\kappa =2 \pi \times 0.05 \,\mathrm{MHz}$, $ \gamma = 2 \pi \times 11 \,\mathrm{MHz}$ and $\omega_\mathrm{rec}=2 \pi \times 6.8 \,\mathrm{kHz}$.}
\label{fig4}
\end{figure}
 In Fig.~\ref{fig4}a), we plot $g^{(2)}(0)$ as a function of laser-cavity detuning $\delta_c$ and the central position $x_0$ of the trap, for parameters  $g_0 = 2 \pi \times 10 \,\mathrm{MHz}$, $\kappa = \gamma =2 \pi \times 0.02 \,\mathrm{MHz}$,  $\Delta=5g_0$ and $\omega_m= 2 \pi \times 0.5  \,\mathrm{MHz}$. It can be seen that this figure captures a combination of the pure J-C plot (Fig.~\ref{fig2}b) and pure optomechanical plot (Fig.~\ref{fig1}d), where the largest degree of antibunching occurs around the anti-node $(x_0=0)$ or in between the node and anti-node, respectively. In particular, the presence of sideband features, and the extended antibunching away from the anti-node are qualitative signatures of motional effects. Below we plot $g^{(2)}(0)$  following the ZPL (red, dashed). 
The region of negligible antibunching, $g^{(2)}(0)\approx 1$, at $k_c x_0 \approx \pm \pi/8$ originates from an exact cancellation of the nonlinearities induced by motion and the two-level nature. 

To better understand the contribution from motion, under certain conditions one can effectively map the J-C model to the optomechanical Hamiltonian. In particular, for large laser-atom detunings $\delta_0 \gg g_0$, the atomic ground state population is approximately one which allows for an effective elimination of the atomic excited state \cite{stefan,ritsch} using the Nakajima-Zwanzig projection operator formalism \cite{naka,zwanzig}.
The Lamb-Dicke regime is given by $\eta_\mathrm{LD} = k_c x_\mathrm{zp} =\sqrt{\omega_\mathrm{rec}/\omega_m} \ll 1$, where the atomic recoil frequency  $\omega_\mathrm{rec}= \hbar k_c^2/(2m_\mathrm{atom})$ relates the resonant wavevector with the atomic mass.
In this regime, the effective optomechanical Hamiltonian (\ref{int}) is reproduced 
by replacing  $g_m \to g_\mathrm{eff}$ with the effective optomechanical coupling strength
$
 g_\mathrm{eff} = g_0^2 \delta_0/(\delta_0^2+ \gamma^2/4) \eta_\mathrm{LD} \sin( 2 k_c x_0)
$
and $\kappa \to  \kappa_\mathrm{eff}$ with the effective cavity linewidth
$
 \kappa_\mathrm{eff} = \kappa + \gamma  g_0^2/(\delta_0^2 + \gamma^2/4) u^2(x_0),
$ broadened by atomic spontaneous emission (see Appendix B). 
Note that $\delta_0 \approx -\Delta$ for $\Delta \gg g_0$ and when the system is driven resonantly on the ZPL.
For small $\eta_\mathrm{LD}$, the nonlinearity arising from motion simply adds to that arising from the two-level nature of the atom, and the energy spectrum reads
\begin{equation}
 E_\mathrm{n,m} \approx m \omega_m + \left(\omega_c - \frac{g_0^2}{\Delta} u^2(x_0) \right)n + \left( \frac{g_0^4}{\Delta^3} u^4(x_0) - \frac{g_\mathrm{eff}^2}{\omega_m} \right)n^2.
\end{equation}
Here, $n$ denotes the number of excitations in the photon-like eigenstate of the J-C model. Thus, the essential ingredients needed to observe a quantum nonlinearity associated with the motion are  $g_\mathrm{eff}^2/\omega_m\gtrsim\kappa_\mathrm{eff}$ and $\omega_m\gtrsim\kappa_\mathrm{eff}$, along with $\eta_\mathrm{LD} < 1$, such that the atomic motion can be linearized. Even for relatively large Lamb-Dicke parameters ($\eta_\mathrm{LD} \approx 0.25$) higher order corrections are minor, as we show in the Appendix C. As the two-level and motional anharmonicities scale with $\Delta^{-3}$ and $\Delta^{-2}$, respectively,  increasing $\Delta$ serves as a way to make two-level antibunching vanish while nonlinear motional effects persist. Furthermore, as the maximum allowed value of $g_\mathrm{eff}$ to retain validity of the effective model is $g_\mathrm{eff}\sim g_0$, one can see that the cavity QED strong coupling condition $g_0\gtrsim \kappa$ naturally enables optomechanical strong coupling. Actually, the more conventional criterion for cavity QED strong coupling, $g_0>\kappa,\gamma$, is not required, as we illustrate next.

To present the realistic possibilities of observing optomechanical blockade, we consider an existing cavity QED setup with trapped $^{40}\mathrm{Ca}^+$-ions \cite{tracy1} with $g_0 = 2 \pi \times 1.4 \,\mathrm{MHz}$, $\kappa =2 \pi \times 0.05 \,\mathrm{MHz}$ and $ \gamma = 2 \pi \times 11 \,\mathrm{MHz}$. Note that without motion, the large spontaneous emission rate $\gamma\gg g_0$ in this particular setup prevents one from observing blockade arising from the Jaynes-Cummings ladder when the atom and cavity are on resonance. Blockade cannot be observed by working off resonance either, as the nonlinearity in the spectrum decreases faster ($\propto \Delta^{-3}$) than the atomic contribution to the decay rate of the cavity ($\propto \Delta^{-2}$). However, optomechanical blockade can be observed as its nonlinearity decreases also as $\Delta^{-2}$. In Fig.~\ref{fig4}b) we plot $g^{(2)}(0)$ as a function of atom position $x_0$ and detuning $\delta_c$, for $\Delta=12g_0$ and $\omega_m = 2 \pi \times 0.1 \,\mathrm{MHz}$, and also for a detuning $\delta_c$ following the ZPL (red, dashed). 
As the maximal two-level anharmonicity $ 2 (g_0^4/\Delta^3) u^4(0) \approx 2 \pi \times 1.6 \,\mathrm{kHz} \ll \kappa_\mathrm{eff}$ is far from being resolved, no photon blockade occurs due to the two-level nature and thus no antibunching can be seen at the anti-nodes. However, the
motional nonlinearity $2 g_\mathrm{eff}^2/\omega_m \approx 2 \pi \times 15 \,\mathrm{kHz}$ is almost an order of magnitude larger and allows a minimum value of $g^{(2)}(0) \approx 0.83$ driving the ZPL  around $k_c x_0\approx \pi/3$.

This value actually represents the optimum that can be observed at this position, scanning over the parameters $\omega_m$ and $\Delta/g_0$ as we illustrate in Fig.~\ref{fig4}c). For lower values of $\Delta$, the sideband resolution is lost owing to the large value of the atomic spontaneous emission rate $\gamma$ and its contribution to the effective cavity linewidth $\kappa_\mathrm{eff}$ ($\kappa_\mathrm{eff}\approx 2 \pi \times 84 \, \mathrm{kHz}$ at the optimized point). On the other hand, for increasing $\omega_m$, the magnitude of the motional nonlinearity $2g_\mathrm{eff}^2/\omega_m$ becomes reduced, while for decreasing $\omega_m$ again sideband resolution is lost. Note as well that the anti-bunching is negligible for any detuning, when the motion is frozen out ($\omega_m \rightarrow \infty$). This dependence of $g^{(2)}(0)$ on $\omega_m$ reveals the pure motional origin of antibunching. Fig.~\ref{fig4}d) shows $g^{(2)}(0)$ as a function of atom position $x_0$ and trap frequency $\omega_m$, for $\Delta=12g_0$ and resonantly driving the ZPL. Here one again sees that the antibunching occurs only between the nodes and anti-nodes, and the tradeoff in $\omega_m$. 

In conclusion, we have shown that cavity QED experiments approaching the strong coupling regime are natural platforms to explore the single-photon, single-phonon strong coupling regime of optomechanics, in the limit that the motional sidebands can be resolved. Since many of those experiments, which allow for the realization of motional nonlinear effects, already exist, we anticipate that such platforms will stimulate much theoretical and experimental work to further explore the generation of non-classical light from motion and its consequences.
\begin{acknowledgments}
 The authors thank H.J. Kimble and P. Rabl for stimulating discussions. Lukas Neumeier acknowledges support from the Ministry of Economy and Competitiveness of Spain through the ``Severo Ochoa'' program for Centres of Excellence in R\&D (SEV-2015-0522), Fundaci\'{o} Privada Cellex, Fundaci\'{o} Privada Mir-Puig, and Generalitat de Catalunya
through the CERCA program. DEC acknowledges support from the Severo Ochoa Programme, Fundacio Privada Cellex, CERCA Programme / Generalitat de Catalunya, ERC Starting Grant FOQAL, MINECO Plan Nacional Grant CANS, MINECO Explora Grant NANOTRAP, and US ONR MURI Grant QOMAND.
TEN acknowledges support from the Austrian Science Fund (FWF) Projects Y951-N36 and F4019-N23.

\end{acknowledgments}

\appendix 

\section{Calculation of second-order photon correlations $g^{(2)}(0)$}
Here, we discuss how to calculate the second-order correlation function $g^{(2)}(0)$ of the transmitted field given a weak coheren state input, such as plotted in Figs. 1c), 1d), 2b) and Fig. 3 of the main text.
Formally, the quantum properties of the transmitted field are encoded in the input-output relation $a_\mathrm{out}(t) = a_\mathrm{in}(t) + \sqrt{\kappa/2} a(t)$. As the external driving field is injected through the other mirror, the input field in the transmitted port is the vacuum state, and thus the second-order correlation function $g^{(2)}(0)= \langle (a_\mathrm{out}^\dag)^2 a_\mathrm{out}^2 \rangle/\langle a_\mathrm{out}^\dag a_\mathrm{out} \rangle^2 =  \langle (a^\dag)^2 a^2 \rangle/\langle a^\dag a \rangle^2 $ depends only on the intra-cavity field.
We numerically calculate the necessary expectation values from the system wavefunction $\ket{\Psi(t)} = \sum_{n,m} c_\mathrm{n,m}(t) \ket{n,m}$ (where $n$ denotes the photon number and $m$ the phonon number), which we truncate for $n_\mathrm{max}>2$ (given that a sufficiently weak input state is unlikely to generate more than two cavity photons), and $m_\mathrm{max}$ depending on convergence. In the case of the pure optomechanical Hamiltonian $H_{\mathrm{op}}$, we solve for the steady-state amplitudes $c_{n,m}$ from the effective Schroedinger equation $\I \ket{\dot{\Psi}(t)} = H_\mathrm{op} \ket{\Psi(t)}$.  Then, $\langle a^\dag a \rangle = \sum_m   |c_{1,m}|^2+2 |c_{2,m}|^2$ and $\langle (a^\dag)^2 a^2 \rangle = \sum_m 2 |c_{2,m}|^2$. Note that we neglect mechanical damping as our true subspace of interest consists of trapped atoms. Formally, the inclusion of cavity dissipation in the effective wavefunction evolution must be supplemented with stochastic quantum jumps \cite{elements}. However, in the weak driving limit $E_0 \rightarrow 0$ that we consider here, the effect of jumps on observables becomes vanishingly small and thus we do not need to explicitly account for them. While we have explicitly discussed the optomechanical Hamiltonian $H_\mathrm{op}$ here, the cases of the Jaynes-Cummings model without motion or Jaynes-Cummings model including motion are solved in an immediately similar fashion.
\section{Derivation of effective optomechanical coupling $g_\mathrm{eff}$ and effective cavity linewidth $\kappa_\mathrm{eff}$}
The full master equation corresponding to the Hamiltonian $H_\mathrm{JC}$ (Eq.~(2) of the main text), where we treat $x_0 \to x$ as a dynamical variable is given by
\begin{equation}\label{fullmodel}
 \dot{\rho} = -\I \left(H_\mathrm{JC}\rho - \rho H^\dag_\mathrm{JC}\right) + \sigma_\mathrm{ge} e^{- \I k_c  x} \rho e^{\I k_c x} \sigma_\mathrm{eg} + \kappa a \rho a^\dag \equiv L\rho.
\end{equation}
The term $ \sigma_\mathrm{ge} e^{- \I k_c  x} \rho e^{\I k_c x} \sigma_\mathrm{eg}$ physically describes quantum jumps corresponding to atomic spontaneous emission, accompanied by a momentum recoil kick $e^{- \I k_c x}$ acting on the atomic motion. For simplicity, we only consider a single direction of spontaneous emission. 

In the limit where the cavity is driven near resonantly and the atom is far-detuned, the atomic excited state can be eliminated to yield an effective optomechanical system involving just the atomic motion and the cavity mode. We will now use the Nakajima-Zwanzig projection operator formalism to eliminate the atomic excited state. We define a set of operators $P,Q$, which project the entire system density matrix 
\begin{equation}\label{ro}
 \rho = |g\rangle \langle g| \rho_{gg} +  |g\rangle \langle e| \rho_{ge}+|e\rangle \langle g| \rho_{eg}+|e\rangle \langle e| \rho_{ee},
\end{equation}
into the subspace spanned by $\ket{g}\bra{g}$ (which we want to project the dynamics into), and its orthogonal  $\id-\ket{g}\bra{g}$.
Here $\rho_{ij} =\langle  i |\rho| j \rangle $ are the reduced density matrices for the reduced Hilbert space, which still contain all other existing degrees of freedom. Thus, we define a projection operator $P$:
\begin{equation}
 P \rho = |g\rangle \langle g| \rho_{gg}
\end{equation}
and its complementary
\begin{equation}
 Q \rho =    |g\rangle \langle e| \rho_{ge}+|e\rangle \langle g| \rho_{eg}+|e\rangle \langle e| \rho_{ee}.
\end{equation}
It is straightforward to show $P^2=P,Q^2=Q, QP=0,P+Q = \id$.
We will now  divide the super operator $L$ up in parts 
according to the way they act on the Hilbert space describing the internal degrees of freedom of the atom:
\begin{equation}
 L =L_o + L_a + L_I + J.
\end{equation}
Here, $L_o=L_m+L_c$ is composed of terms that do not act on the internal degrees of freedom, with $L_m$ and $L_c$ describing respectively the trapped atomic motion and the bare dynamics of the driven cavity mode:
\begin{equation}\label{Lm}
 L_m \rho= -\I [\omega_m b^\dag b,\rho]
\end{equation}
\begin{align}\label{Lc}
L_c\rho & \nonumber=  \I \delta_c [ a^\dag a, \rho ] - \I \sqrt{\kappa/2} E_0  [(a+a^\dag),\rho ] \\& -  \frac{\kappa}{2} \left( a^\dag a \rho + \rho a^\dag a - 2 a \rho a^\dag \right).
\end{align}
The super operator
\begin{equation}\label{La}
 L_a \rho =  \I  \delta [ \sigma_{ee} ,\rho] - \frac{\gamma}{2} \{ \sigma_{ee}, \rho \}
\end{equation}
acts on $|e\rangle \langle g|,|g\rangle \langle e|,|e\rangle \langle e|$ (the subspace spanned by $Q$) and just multiplies those terms by a c-number. It describes evolution and damping of the excited internal state of the atom.
\begin{equation}\label{Li}
 L_I \rho = - \I [g(x)(\sigma_\mathrm{eg}a + \sigma_\mathrm{ge}a^\dag),\rho]
\end{equation}
acts on all the states and all Hilbert spaces, describing the interaction of the atom with the cavity field and
\begin{equation}\label{J}
 J\rho = \gamma  \sigma_\mathrm{ge} e^{- \I k_c x} \rho e^{\I k_c x} \sigma_\mathrm{eg}
\end{equation}
describes the spontaneous jump of the excited state of the atom into its ground state accompanied by a momentum recoil.
We define $v=P\rho$ and $w=Q\rho$ and insert $P+Q = \id$ into Eq.~(\ref{fullmodel}):
\begin{equation}
 \dot{v} = P\dot{\rho} = P L \rho = P L P \rho+ P L Q \rho 
\end{equation}
After identifying all vanishing terms, we obtain:
\begin{equation}\label{dd1}
 \dot{v} = L_o v + P (J + L_I) w 
\end{equation}
and
\begin{equation}\label{ww1}
 \dot{w} = QL_Iv + Q (L_o+ L_a + L_I) w. 
\end{equation}
Since $w$ describes the evolution of the fluctuations out of the subspace of interest, the term $L_o w =(L_m + L_c)w$ describes the free evolution of motion and of the cavity mode during one of these fluctuations. As the timescale of these fluctuations is set by $\delta_0$ and $\gamma$ and we assume that either $\delta_0$ or $\gamma$ is much larger than both $\omega_m$ and $\kappa$, we can neglect the time evolution of motion and cavity during one of these fluctuations by approximating $L_o w \approx 0$ in Eq.~(\ref{ww1}). Then the general solution to this equation reads:
\begin{align}
 w(t) & \nonumber = \int_0^t d\tau e^{Q(L_o +L_a)(t-\tau)} Q L_I w(\tau) \\& + \int_0^t d\tau e^{Q(L_o+L_a)(t-\tau)} Q L_I v(\tau)
\end{align}
where we set $w(0)=0$ as the initial condition.
Now we plug this equation twice into Eq.~(\ref{dd1}) (iterative) in order to catch a term of the order $J L_I^2$ and include the process of spontaneous emission:
\begin{align}
 \dot{v}(t)  = & \nonumber L_o v + P(J +L_I) \int_0^t d\tau e^{Q(L_o +L_a)(t-\tau)} Q L_I v(\tau) \\ & + P(J +L_I)  \int_0^t d\tau e^{Q(L_o +L_a)(t-\tau)} Q L_I \nonumber \\& 
 \times \int_0^\tau d\tau' e^{Q(L_o +L_a)(t-\tau')} Q L_I v(\tau')
\end{align}
where we neglected the term proportional to $w(\tau')$ since it produces only terms $\propto L_I^3$ or higher. After identifying vanishing terms, we are left with:
\begin{align}\label{las1t}
 & \dot{v}(t)  \nonumber = L_o v+ PL_I \int_0^t d\tau e^{(L_o +L_a)(t-\tau)}  L_I v(\tau) \\ & + PJ \int_0^t d\tau e^{(L_o +L_a)(t-\tau)}  L_I \int_0^\tau d\tau' e^{(L_o +L_a)(t-\tau')}  L_I v(\tau').
\end{align}
After extending the lower integral borders to $-\infty$ (Markov approximation) and evaluating the integrals, we obtain the effective optomechanical master equation:
\begin{equation}\label{mastaop}
 \dot{\rho} = -\I \left[H_\mathrm{om},\rho \right] + L_\mathrm{om}\rho, 
\end{equation}
with an effective optomechanical Hamiltonian
\begin{equation}\label{gfn}
 H_\mathrm{om}= \omega_m b^\dag b - \Delta_c(x) a^\dag a + \sqrt{\kappa/2} E_0 (a + a^\dag).
\end{equation}
The position dependent cavity-laser detuning is given by
\begin{equation}\label{cav}
 \Delta_c(x)=  \delta_c -  \frac{g_0^2 \delta_0}{\delta_0^2 + \frac{\gamma^2}{4}} u^2(x). 
\end{equation}
Expanding $ \Delta_c(x)$ around $x_0$ to linear order and replacing $x$ with phonon operators $b$ and $b^\dag$ yields $ \Delta_c(x) \approx \Delta_c(x_0) - g_\mathrm{eff}(b+b^\dag)$ with $g_\mathrm{eff}$ of the main text.
The system losses are given by the effective Liouvillian
\begin{align}\label{Lc}
L_\mathrm{om}\rho & \nonumber =- \frac{\kappa}{2} \left( a^\dag a \rho + \rho a^\dag a - 2 a \rho a^\dag \right)  \\ & -  \frac{\gamma}{2}\frac{ g_0^2}{ \delta_0^2 + \frac{\gamma^2}{4}} \left(u^2(x) a^\dag a \rho  +  \rho a^\dag a u^2(x) \right) \nonumber \\&
  +\frac{\gamma}{2}\frac{ g_0^2}{ \delta_0^2 + \frac{\gamma^2}{4}}
 \left( 2 a u(x) e^{- \I k_c x} \rho e^{\I k_c x} u(x) a^\dag\right), 
\end{align}
which describes the broadening of the cavity linewidth due to atomic spontaneous emission,  
\begin{equation}\label{ka}
 \kappa(x) = \kappa + \gamma  \frac{g_0^2}{\delta_0^2+ \frac{\gamma^2}{4}} u^2(x). 
\end{equation}
Averaging with the atomic wavefunction located at $x_0$ yields $\kappa_\mathrm{eff}$ of the main text.

\begin{figure}
     \centering
\includegraphics[width=0.5\textwidth]{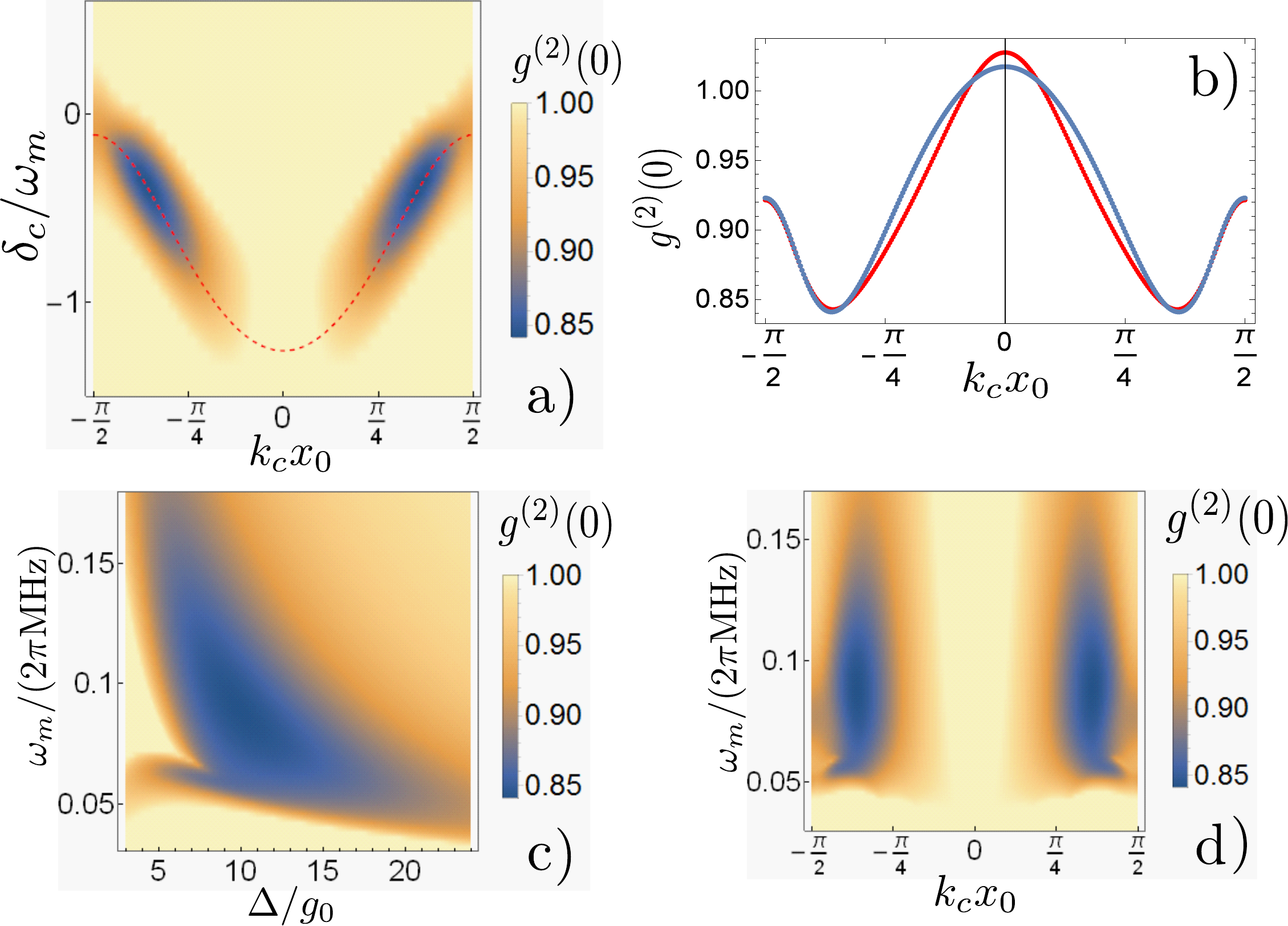}
\caption{J-C model with motion expanding $u(x)$ until quadratic order. \textbf{a)} $g^{(2)}(0)$ of the transmitted field versus trapping position $x_0$ and detuning from the empty cavity $\delta_c=\omega_L-\omega_c$, for detunings near the photonic eigenstate and by using the parameters for a realistic cavity QED experiment given below. In this figure, we choose an atom-cavity detuning $\Delta=10g_0$ and atomic trap frequency $\omega_m = 2\pi \times 0.09 \,\mathrm{MHz}$, which produces the minimum possible $g^{(2)}(0)$ including quadratic order corrections.  
\textbf{b)} Following the ZPL of a) (red, dashed). W compare  $g^{(2)}(0)$ calculated with only linear displacements (red) in Hamiltonian Eq.~(2) of the main text with $g^{(2)}(0)$ calculated by also including terms of quadratic order (blue).   
\textbf{c)} $g^{(2)}(0)$  as  a function of atom-cavity detuning $\Delta$ and trapping frequency $\omega_m$ including terms of quadratic order. order. Here, the atomic position is fixed at $k_c x_0= 1.15$.
\textbf{d)} $g^{(2)}(0)$  as  a function of trapping position $x_0$ and trapping frequency $\omega_m$ for $\Delta = 10g_0$ including terms of quadratic order. As in the main text, we choose parameters of an existing cavity QED experiment with trapped $^{40}\mathrm{Ca}^+$-ions: $g_0 = 2 \pi \times 1.4 \,\mathrm{MHz}$, $\kappa =2 \pi \times 0.05 \,\mathrm{MHz}$, $ \gamma = 2 \pi \times 11 \,\mathrm{MHz}$ and recoil frequency  $\omega_\mathrm{rec}=2 \pi \times 6.8 \,\mathrm{kHz}$.}
\label{figSM1}
\end{figure}
\section{Beyond the Lamb-Dicke regime: Including quadratic-order terms in displacement}
In order to show that the strong coupling regime of optomechanics can already be observed by an existing experiment, we plotted  $g^{(2)}(0)$  as a function of $x_0$ in Fig. 3b) in the main text. In this calculation, we linearized the cavity mode profile $u(x)$ in Hamiltonian Eq.~(2) of the main text around the trapping position $x_0$: $u(x) \approx u(x_0) + u'(x_0) k_c (x-x_0) $,
 which is strictly only valid in the Lamb-Dicke regime $\eta_\mathrm{LD} = k_c x_\mathrm{zp} = \sqrt{\omega_\mathrm{rec}/\omega_m} \ll 1$.
However, in order to produce Fig. 3 of the main text, we used a trapping frequency of $\omega_m = 2 \pi \times 0.1 \,\mathrm{MHz}$. With the recoil frequency of  $^{40}\mathrm{Ca}^+$-ions this corresponds to $\eta_\mathrm{LD} \approx 0.26$.

To ensure that the results are not significantly affected by this relatively large Lamb-Dicke parameter, we will now include the next order term $u(x) \approx u(x_0) +  u'(x_0) k_c (x-x_0) +(1/2) u''(x_0) (x-x_0)^2$.  In Fig. \ref{figSM1}a), we plot the adjusted $g^{(2)}(0)$ as a function of atom position $x_0$ and detuning $\delta_c$. Here we choose $\Delta=10g_0$ and $\omega_m = 2 \pi \times 0.09 \,\mathrm{MHz}$ in order to minimize $g^{(2)}(0)$ including quadratic order corrections.
Fig. \ref{figSM1}b) shows $g^{(2)}(0)$ as a function of atom position $x_0$ following the ZPL of a) (blue). In red we plot $g^{(2)}(0)$, where $u(x)$ has only been expanded until linear order for the same parameters. We observe a reasonable match and conclude that linearizing motion on the Hamiltonian level at least qualitatively fully captures the relevant physics even for relatively large $\eta_\mathrm{LD}$. For completeness, we plot $g^{(2)}(0)$ as a function of $\omega_m$ and $\Delta$ in Fig. S1c) for a fixed atomic position $k_c x_0=1.15$, and in Fig. S1d) we plot $g^{(2)}(0)$ as a function of trapping position $x_0$ and trap frequency $\omega_m$ for $\Delta = 10 g_0$.

\end{document}